\documentclass[12pt,twoside]{article}
\usepackage{fleqn,espcrc1}
\usepackage{graphicx}
\title{Space-time structure of a bound nucleon}

\author{A.V. Molochkov
\address{BLTP, Joint Institute for Nuclear Research, 141980, Dubna, Russia}}
       
\begin{document}

\maketitle

\begin{abstract}
Distortions of bound nucleon space-time structure, measured
in deep inelastic scattering off nuclei, are discussed from view-point of 
a theoretical approach based on the Bethe-Salpeter formalism.
It is shown that modification of the structure function $F_2^{\rm N}$ 
results from relative time dependence in Green functions of 
bound nucleons. 
The modification plays fundamental role in analysis of deep inelastic 
scattering experiments and allows one to obtain new information 
about nucleon 
structure. 
\end{abstract}

\section{INTRODUCTION}

Decades of matter structure studying has shown that 
complete information about nucleon structure cannot be obtained 
from free nucleon data only. First of all, it is connected with absence
of a stable enough free neutron target. The attempts to solve this problem
were based on utilization of nuclear targets. However, small nuclear
effects, which were supposed to be negligible, led to the 
qualitatively different
results for the structure function of a nucleon bound in a deuteron and 
in heavy nuclei~\cite{EMC}.
This phenomenon reflects untrivial difference between nuclear 
and nucleon structure at internucleon distances and its evolution 
with atomic number $A$. 
Nature of the effect was analyzed in 
numerous models which were produced 
since it was discovered (comprehensive reviews
of the models can be found in~\cite{reviews}). 
Summing up the basic qualitative pictures, proposed  in the models, 
one can conclude that the difference between nucleon and nuclear 
structure functions should originate from 
properties of nucleon structure and structure of nucleon-nucleon interaction.  
Thus, from the one hand the nuclear effects should be 
defined by properties the $n$-nucleon Green functions ($n=1,2\dots A$), 
from the
other hand detailed information about {\it nuclear} effects can provide new
information about {\it nucleon} structure.
In this paper the general properties of relativistic bound states, which 
presumably lead
to the EMC-effect, are discussed. 
It is shown that the  relativistic consideration allows one to establish new
regularities in nuclear structure which are important
in the construction of the nucleon parton distributions.  

\section{SPACE-TIME STRUCTURE OF RELATIVISTIC BOUND STATES}

The amplitude of deep inelastic scattering is defined by the imaginary part of 
the forward Compton amplitude:
\begin{eqnarray}
W^{A}_{\mu \nu}(P,q) = {\rm Im}_{q_0}i \int d^4x e^{ i q x } 
\langle A,P|  {\rm T}
\left(J_\mu \left(x \right) J_\nu \left(0\right) \right) |A,P
\rangle \nonumber
\end{eqnarray}
In the framework of the Bethe-Salpeter formalism the amplitude is expressed 
in terms of solutions of the homogeneous Bethe-Salpeter equation~\cite{BS}
\begin{equation}
{\chi}_{\alpha, P}^A({\cal X})= 
\int d{\cal Z} d{\cal Z}^\prime
S_{(n)}({\cal X}, {\cal Z}) \overline G_{2n}({\cal Z}, {\cal Z}^\prime) 
{\chi}_{\alpha, P}^A({\cal Z}^\prime)
\label{BS}\end{equation} 
and the Mandelstam vertex
${\overline G_{2(n+1)}}_{\mu\nu}({\cal Z}, x, {\cal Z}^\prime)$~\cite{mandel}: 
\begin{equation}
\langle A,P|{\rm T}(J_{\mu}\left(x)J_{\nu} (0)\right)|A,P\rangle
=\int d{\cal Z} d{\cal Z}^\prime
\bar{\chi}_{\alpha, P}^A({\cal Z})
{\overline G_{2(n+1)}}_{\mu\nu}({\cal Z}, x, {\cal Z}^\prime)
{\chi}_{\alpha, P}^A({\cal Z}^\prime).
\end{equation}
The calligraphic letters denote a set of nucleons positions in the 
four-dimensional 
space -- ${\cal Z}=z_1,\dots z_n$.
Relative positions of the nucleons are defined by four-dimensional 
intervals -- 
$
r_i=\sum_{j}^{n} z_j/n-z_i.
$
This is the most unusual feature of relativistic bound states 
that nucleons inside it 
are not only divided by three-dimensional intervals but also {\it shifted
in time}. The shift is defined by 
the zero component of the interval -- {\it relative time}  $\tau_i={r_i}_0$. 

If one considers nucleons as three-dimensional objects then the shift 
in time looks like unphysical 
feature~\cite{quasi}. One should assume in this case that the shift 
does not affect on observables, and  
different values of the variables $\tau_i$ in Eq.~(\ref{BS}) should 
lead to equivalent quasi-potential approaches~\cite{quasi}. 
Recently it was shown, however, that the approaches are not equivalent from
the point of view of relativistic covariance~\cite{Pascalutsa}. 
In general, the covariance can be 
kept only in equal-time approaches where $\tau_i=0$. 
This contradiction points to existence of observable effects which could
result from the shift in time of bound nucleons. Existence 
of such effects can be natural within the hypothesis 
that bound nucleons are shifted in time four-dimensional objects. 

\begin{figure}[t]
\includegraphics[width=35pc,height=11pc]{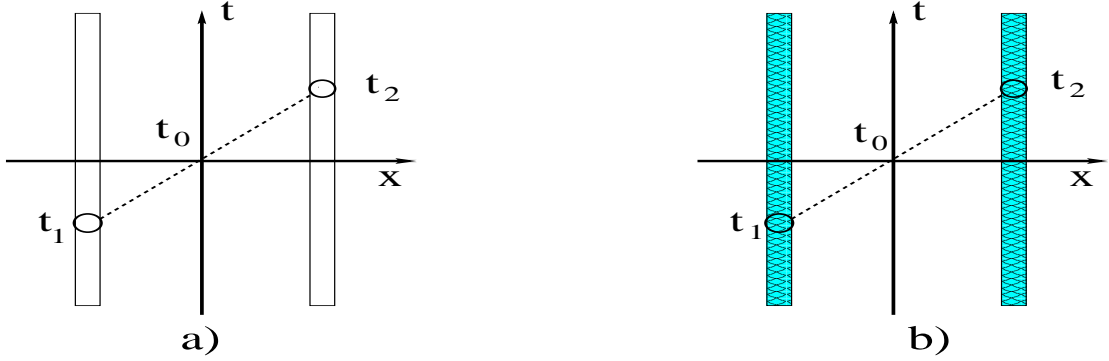}
\caption{\label{async} Schematic presentation of the 
space-time diagram for the two-nucleon bound state 
for the cases when the nucleon has a static (a)
and dynamic (b) structure. }
\end{figure}

The space-time diagram for the bound state of two nucleons, which are
considered to be four-dimensional objects, can be represented as
it is shown in Figure~\ref{async}.
The Fermi motion is disregarded for simplicity. 
The static structure shown in diagram (a) is homogeneous in time, while 
diagram (b) represents an inhomogeneous in time bound nucleon structure. 
If one makes an instant flash of the bound state at the moment $t_0$ 
one finds that 
the structure is defined by one nucleon at the moment $t_1$ in the future 
and by another one at the moment $t_2$ in the past. 
If the nucleon had the static structure 
then the three points $t_0, t_1, t_2$ 
in Figure~\ref{async}(a) would be equivalent to each other and 
no effect would be observed from the shift in time.
In another case,
the points in Figure~\ref{async}(b)
are not equivalent to each other
due to the evolution of the nucleon structure in time. The shift
in time would then show up in modifications of the observables.  
From the stability of bound nucleons one can infer that the evolution
of the nucleon structure in time has a periodic character.
The period $\theta$
should be comparable to the mean value
of relative time absolute  $\theta \sim |\tau|$. If 
$|\tau| \gg \theta$ or $|\tau| \ll \theta$ then no effect from the shift 
in time of bound nucleons can be measured. 

The established relation between
the structure of the nucleon in  the future and the
measurements at present looks as if  
the causality principle is  violated. 
However, the quantum nature of the considered objects does not allow
one to perform an instant mea\-su\-re\-ment. Instead one have to consider
the contribution of 
nucleon states averaged between the time moments $t_1$ and $t_2$.
 This suggests a qualitative interpretation of the modification of the
 bound nucleon structure.
One can estimate from 
the uncertainty relation the size of the area in which the nucleon
can be localized  in 4-space.
 The radius $\rho$ of the area,  ``nucleon localization radius'',
can be related to the nucleon mass as follows:
$
\rho^2=\Delta t^2 - \Delta x^2\sim 1/m_{\rm N}^2.
$
The shift in time results in additional uncertainty  $|\tau|=|t_2-t_1|$ 
in definition of the time moment of the measurement:
$
\widetilde\rho^2=( \Delta t+|\tau| )^2 - \Delta x^2\sim 1/{m^{*}_{\rm N}}^2.
$
Since $|\tau| > 0$ by definition, the localization radius 
becomes larger $\widetilde\rho > \rho $ and 
effective mass of the nucleon becomes smaller $m^* < m$.
This is similar to the hypothesis of the $x$-rescaling model~\cite{xresc},
where the Bjorken variable $x$ of a bound nucleon is rescaled 
$
\tilde x=x/(1-\epsilon/m_{\rm N})
$ 
due to the
shift of nucleon mass by the rescaling parameter $\epsilon$: 
$
m^*=m-\epsilon
$. 
The $x$-rescaling model
gives qualitative description of the  deviation of nuclear 
to nucleon structure 
function ratio from unity (EMC -- effect).
Thus we can conclude that the shift in time manifests itself in
observables, the EMC effect being one of the examples.

This conclusion has been justified in general terms in the
publications~\cite{nucphys,phlt} where the nuclear structure 
functions were calculated in the framework of the covariant approach based
on the Bethe-Salpeter formalism.
In the framework of this approach the relative time 
dependence has been consistently taken into account 
with the help of
analytical properties of nucleon Green functions. 
The developed approach allows one to consider 
the theory with equal-time 
bound nucleons as the leading approximation in
description of relativistic bound states, where the relativistic 
corrections, connected with the relative time 
dependence, are disregarded. In order to take the corrections into account,  
one has to keep the dependence on the relative time through 
the entire chain of calculation of observables.
As a result, the 
nuclear structure function has been obtained in the form:
\begin{equation}
F_2^{A}(x_{A})=
\int\frac{d^3k}{(2\pi)^3} \sum\limits_{a,a^\prime}^{A-1}
\left[\frac{E_{a}-{k_3}}{E_{
a}}F_2^{a}(x_{a})+ \frac{\Delta^{A}_{a,a^\prime}}{E_{a}}
x_{a}\frac{d F_2^{a}(x_{a})} {dx_{a}} \right]
\Phi_{a,a^\prime}^{A}({\bf k})^2,\label{f2a}
\end{equation}
where the Bjorken variables for the nucleus $A$ and the nuclear 
fragment $a$ are
introduced in the form  
$x_{\mbox{\tiny\it A}}=Q^2/(2P_A\cdot q)$
and 
$x_{a}=Q^2/(2p_{a}\cdot q)$.
The function $\Phi_{a,a^\prime}^{A}({\bf k})^2$ defines distribution of the 
nuclear fragment $a$ ($a={\rm N, D, ^3He, \dots} $) 
in field of the spectator system $a^\prime$
($a^\prime = {\rm N, NN, D, DN, ^3He, \dots} $). 
The energy of the nuclear fragment 
is denoted as
$E_a=\sqrt{M^2_a+{\bf k}^2}$, $M_a$ is mass of the fragment,
the coefficients   
$\Delta^A_{\rm a,a^\prime}=-M_A+E_{\rm a}+E_{a^\prime}$ can be interpreted
as the removal energy of the corresponding nuclear fragment. 
The term with derivative  of the nucleon structure function in Eq.(\ref{f2a}) 
results from
the shift in time of bound nucleons~\cite{phlt}. 
It is clear that the result can be rewritten in the form
 which one obtains within the 
$x$-rescaling model:
\begin{eqnarray}
F_2^{A}(x_{A})=\int dy d\epsilon 
F_2^{\rm N}\left(\frac{x_{A}}
{y-\epsilon/M_{A}}\right)
\int \frac{d^3k}{(2\pi)^3} 
\frac{my}{E_{\rm N}}
\delta\left(y-\frac{E_{\rm N}-k_3}{m}\right)
\sum\limits_{\rm a^\prime}\Phi^{A}_{\rm N,a^\prime}({\bf k})
\delta\left(\epsilon-\Delta^A_{{\rm N},a^\prime}\right).
\nonumber\end{eqnarray} 

The term 
with derivative in Eq.(\ref{f2a}) leads to the depletion of 
the deuteron to nucleon structure 
functions ratio from unity~\cite{nucphys} in numerical calculations. 
The evaluated ratio $F_2^A(x)/F_2^{\rm D}(x)$ is in good
agreement with the data available for $A = 4$. When evaluated at
$A = 3$, the ratio offers the prediction for the experiments
with $\rm ^3He$, $\rm ^3H$ and D targets.
 On the basis on this prediction the two-stage conception
of $A$-dependence for the nuclear structure functions was proposed. 
At the first stage ($1\le A \le 4$) the pattern of the $F_2^A/F_2^{\rm N}$
changes due to the relative time effects. At the second stage the pattern does
not change but the amplitude of ratio oscillation around unity grows due to 
nuclear density evolution~\cite{sm95}. The calculations based on this 
conception give good
description of heavy nuclear data for the ratio~\cite{phlt}. 

\section{CONCLUSION}

To sum up the discussion presented in this paper, the  
following conclusions can be made.
The proposed qualitative picture shows that the shift in time 
of bound nucleons exposes dynamical properties of a nucleon
structure. 
Both the clear connection between the relativistic picture 
and the $x$-rescaling model and successful description of 
existent EMC effect data 
proves that the EMC effect is manifestation of   
the shift in time of the bound nucleons. 
The relativistic calculations show that the observable effects of
the shift in time     
are defined by the derivative with respect to $x$ of the nucleon structure
functions, which, therefore, reflect the dynamical properties of 
a nucleon structure. Thus, precision  data on 
nuclear to deuteron structure functions ratio 
will provide information about  
the derivative, what is important in construction of nucleon 
parton distributions.

I would like to thank  
A.M.~Baldin, V.V.~Burov, C.~Ciofi degli Atti,
V.A.~Nikolaev and G.I.~Smirnov for fruitful discussions. 
I am grateful to organizers of the meeting for warm hospitality and 
support.

\end{document}